\documentclass[sigconf,nonacm]{acmart}

\usepackage{csquotes}
\usepackage{svg}
\svgsetup{inkscapelatex=false}

\setcopyright{none}
\renewcommand\footnotetextcopyrightpermission[1]{}
\acmConference[JCDL '23]{ACM/IEEE Joint Conference on Digital Libraries}{June 26--30,
  2023}{Santa Fe, NM}
\begin{document}

\title{Visualizing Relation Between (De)Motivating Topics and Public Stance toward {COVID-19} Vaccine}

\author{Ashiqur Rahman}
\email{ashiqur.r@niu.edu}
\orcid{0000-0003-3290-2474}
\affiliation{%
  \institution{Northern Illinois University}
  \streetaddress{1425 W. Lincoln Hwy.}
  \city{DeKalb}
  \state{Illinois}
  \country{USA}
  \postcode{60115-2828}
}

\author{Hamed Alhoori}
\email{alhoori@niu.edu}
\orcid{0000-0002-4733-6586}
\affiliation{%
  \institution{Northern Illinois University}
  \streetaddress{1425 W. Lincoln Hwy.}
  \city{DeKalb}
  \state{Illinois}
  \country{USA}
  \postcode{60115-2828}
}


\begin{abstract}
  While social media plays a vital role in communication nowadays, misinformation and trolls can easily take over the conversation and steer public opinion on these platforms. We saw the effect of misinformation during the {COVID-19} pandemic when public health officials faced significant push-back while trying to motivate the public to vaccinate. To tackle the current and any future threats in emergencies and motivate the public towards a common goal, it is essential to understand how public motivation shifts and which topics resonate among the general population. In this study, we proposed an interactive visualization tool to inspect and analyze the topics that resonated among Twitter-sphere during the {COVID-19} pandemic and understand the key factors that shifted public stance for vaccination. This tool can easily be generalized for any scenario for visual analysis and to increase the transparency of social media data for researchers and the general population alike.
\end{abstract}



\maketitle

\section{Introduction}
Misinformation and anti-vaccine propaganda during the COVID-19 pandemic spread at an unprecedented level, making it difficult for healthcare organizations to reach their goals for vaccine coverage \cite{Depoux2020-ay, Aschwanden2021-mj}. It is essential to understand the information resonating on social media and playing an active role in (de)motivating the public \cite{Abbas2020-bg, Hart2020-pu}. While public confidence in the scientific community is decreasing and trust in the vaccine development process is in question  \cite{Finney-Rutten2021-al, Morris2021-sq}, understanding the information producing these doubts can be critical to motivating people toward vaccination.


In this study, we took a visualization\footnote{The visualization tool can be accessed from this URL: \url{https://ashiqur-rony.github.io/visualize-covid-stance/}} approach to understand the topics spreading on Twitter and (de)motivating the public. Through the visual exploration \cite{sun2022,shaikh2022toward}, we noticed more demotivating topics than motivating ones, but the public stance is overall in favor of {COVID-19} vaccination. One possible explanation is that, even when tweeting with good intentions, some language used in those tweets is backfiring and demotivating people from vaccination. Identifying these topics can help healthcare organizations steer public opinion and encourage the general public to vaccinate.

\section{Methods}

\begin{figure}[htb]
    \centering
    \includegraphics[width=1.0\columnwidth]{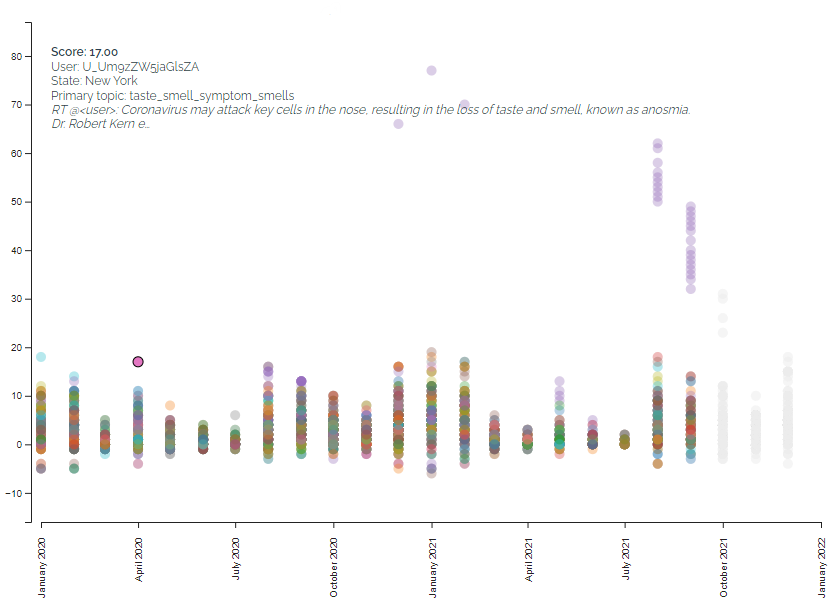}
    \caption{Visualizing the details of a single tweet from the timeline.}
    \Description{Visualizing the details of a single tweet from the timeline.}
    \label{fig:teaser}
\end{figure}

\subsection{Preparing the Datasets}

To create the visualization, we generated a dataset containing tweets related to {COVID-19} vaccine, along with the tweet date, topic of the tweet, stance towards vaccination, and whether the tweet is motivating or demotivating towards vaccination. We prepared the dataset so that all user has at least 20 tweets in the whole dataset. Then we created a \textit{cumulative stance score} for the tweets by adding the scores for each user over time. For this purpose, we considered the stance in favor of vaccination as +1, against as -1, and unrelated as 0, so that when someone is continuously in favor, their score increases over time and vice versa. We chose this for better exploration of public stance over time. If someone changes their stance, we can quickly isolate them and examine their tweets. We also created a \textit{topic frequency} feature containing the number of tweets on a topic each month.

\subsection{Interactive Visualization Tool}

We developed a multiple-view visualization tool to explore the topics and stance from the COVID-19 vaccine-related datasets prepared above. Figure \ref{fig:viz_tool} shows a screenshot from the tool. We have explained each of the views present in the tool below.

\subsubsection{Visualizing Topics}

We used bubble charts to display the topics for a selected month. The horizontal axis represents the topics, while the vertical axis represents the topic frequency calculated in the dataset. The bubbles' size represents the topic's prominence at the selected time segment. We used different colors to isolate the topics in the visualization better. We used an animated hover effect to highlight the tweets in the stance visualization related to topics. While visualizing the topics, we left the generic topic, which contained the stopwords and pronouns out to avoid skewing.

\subsubsection{Visualizing Stance}

We used bubble charts to visualize the cumulative stance over time as displayed in Figrure \ref{fig:teaser}. Each bubble in the visualization represents one tweet. The color of the bubble differentiates the users. User selection of month can highlight the cumulative stance up to that month, and the rest is grayed out. Users can also choose to highlight a single user in the whole visualization. An animated hover effect on a bubble shows details about that tweet, including the cumulative stance score, location, primary topic, and the tweet text.

\subsubsection{User Controls}

\begin{figure}[htb]
    \centering
    \frame{\includegraphics[width=1.0\columnwidth]{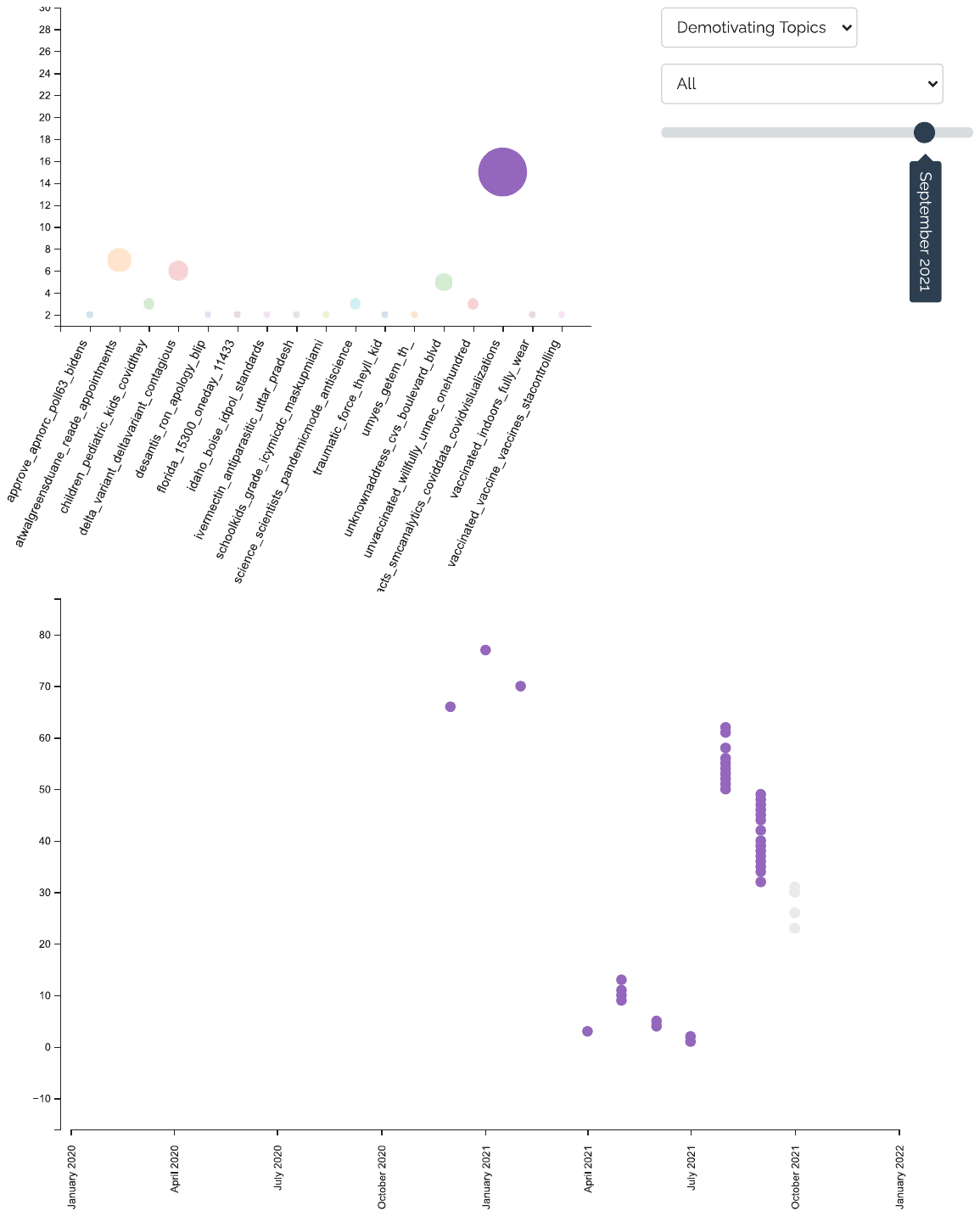}}
    \caption{Highlighting the tweets for one topic by mouse hover.}
    \label{fig:viz_tool}
\end{figure}

There are several user controls present in the visualization. A user can choose to highlight a certain month, user, topic, or tweets. There are two dropdowns and a slider at the top right of the visualization (or in the middle of two visualizations on mobile devices) as displayed in Figure \ref{fig:viz_tool}. Using these controls, the user can interact with and modify the visualizations on the screen. The first dropdown lists the available data sources. The user can select either \textit{motivating} or \textit{demotivating} tweets to populate the visualizations. From the second dropdown, the user can select the author name to highlight tweets from that author in the stance visualization. Using the range slider, the user can select the month to display in the topic visualization and highlight the cumulative stance up to that month. All the controls have animation effects attached to them while changing the visualizations. There are also mouse controls attached to both topic and stance visualization. When the user hovers the mouse pointer on a topic, brushing techniques highlight that topic from the topic visualization at the top and highlight only the tweets using that topic from the stance visualization at the bottom, as displayed in Figure \ref{fig:viz_tool}. The user can also make the selection from the other end. Hovering on a bubble from the stance visualization at the bottom highlights a single tweet, displaying the cumulative stance score, location, primary topic, and tweet text from that tweet text in the top-left corner. The primary topic of the selected tweet is also highlighted in the topic visualization at the top.

\section{Conclusion}

By analyzing (de)motivating topics on Twitter and user stances regarding {COVID-19} vaccine, we noticed that demotivating tweets are more frequent than motivating ones. We also strangely noticed that most demotivating tweets had a favorable stance toward vaccination. This observation led us to believe that while tweeting with good intentions, political polarization, the confirmation bubble, or other factors are demotivating the public about vaccination. This finding opens up new research areas to explore. We also noticed that religion and politics are prominent in the demotivating tweets, while schools, education, and statistical facts are more evident in the motivating tweets. The prominent topics also changed after December 2020, which warrants further study to understand whether the election year played a vital role in shaping public opinion.

We believe this is an important research area that merits future expansion to understand and tackle (de)motivation and public understanding of science to restore public trust in the scientific method. The methods derived from this study can also be extended for any future crisis to tackle the misinformation campaign.


\bibliographystyle{ACM-Reference-Format}
\bibliography{main}

\end{document}